\documentclass[11pt,twoside]{book} 
\usepackage{asp2008s}
\usepackage{times}
\usepackage{lscape}
\usepackage{epsf}

\usepackage{graphicx}
\usepackage{amssymb}
\usepackage{longtable}
\usepackage[figuresright]{rotating}
\usepackage{multirow}
\usepackage[hyperindex,breaklinks]{hyperref}

\newcommand{\simless}{\mathbin{\lower 3pt\hbox {$\rlap{\raise 5pt\hbox{$\char'074$}}\mathchar"7218$}}}

\mainmatter

\newlength{\deftabcolsep}
\begin{document}

\title{The Corona Australis Star Forming Region}   
\author{Ralph Neuh\"auser}
\affil{Astrophysikalisches Institut und Universit\"ats-Sternwarte,
Schillerg\"a{\ss}chen 2-3, D-07745 Jena, Germany}
\author{Jan Forbrich}
\affil{Harvard-Smithsonian Center for Astrophysics, 60 Garden Street MS 72, Cambridge, MA 02138, USA}

\begin{abstract}
At a distance of about 130~pc, the Corona Australis molecular cloud
complex is one of the nearest regions with ongoing and/or recent
star formation. It is a region with highly variable extinction of up
to $A_{V} \sim 45$ mag, containing, at its core, the \textsl{Coronet}
protostar cluster. There are now 55 known optically detected members,
starting at late B spectral types. At the opposite end of the mass
spectrum, there are two confirmed brown dwarf members and seven more
candidate brown dwarfs. The CrA region has been most widely surveyed
at infrared wavelengths, in X-rays, and in the millimeter continuum,
while follow-up observations from centimeter radio to X-rays have
focused on the \textsl{Coronet} cluster.
\end{abstract}

\section{Introduction: The Corona Australis Cloud}

\begin{figure*}
\includegraphics[width=\linewidth]{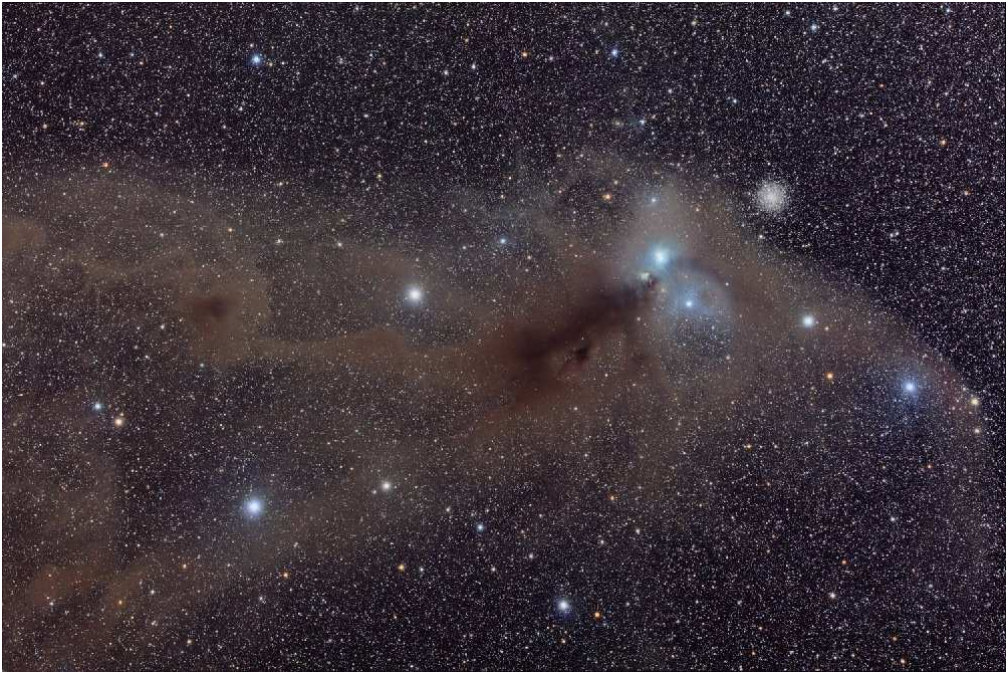}
\caption{Stars and dust in CrA. Credit: Loke Kun Tan (StarryScapes).}
\label{lokekuntan}
\end{figure*}

{\em Corona Australis (southern crown)}, abbreviated {\em CrA},
is one of the 48 ancient constellations
listed by Ptolemy, who called it {\em Stephanos Notios} (in Greek).
The latin translation was {\em corona notia, austrina, australis, meridiana,}
or {\em meridionalis}, from which \citet{all36} picked {\em corona australis};
see also \citet{bot80} for a discussion of the name. For an overview, see Fig.~\ref{lokekuntan}.

\subsection{The Discovery}
\label{sec_disc}

Given the convention for naming variable stars, R CrA was the
first variable star noticed in the CrA constellation.
Marth from Malta and Schmidt from Athens
(quoted in \citealp{rey16}) were the first
to notice the variability of the nebula NGC 6729 in CrA;
Schmidt also discovered the variable stars R, S, and T CrA
with the nebula attached to R CrA.
The variability of stars and nebula were confirmed by
Innes at Cape Observatory (quoted in \citealp{rey16}) and \citet{kno16};
the latter gave $\Delta m$ = 1.4 mag for the variability of R CrA
in images taken from 21 Aug 1911 to 23 May 1915 at Helwan Observatory (Egypt).
\citet{rey16} also noticed that the behaviour of R CrA (variability
and environment) is similar to that of the star T Tau.
\citet{luy27} found the three bright variables UZ CrA, VW CrA, and VX CrA.
Then, \citet{vge32,vge33} found more than 100 variable
stars in or near CrA, based on plates taken at the
Union Observatory in Johannesburg, mostly by himself and Hertzsprung,
a few of them young stars.

\citet{her60} estimated the age of the two early-type emission-line stars
R CrA (Ae) and T CrA (F0e) as their main sequence contraction time
($10^{5}$ to $10^{7}$ years) and
concluded that they are young, and that therefore also
the associated nebula and the other stars should be young.
\citet{kna73} found infrared (IR) excess
in some of the variable stars in CrA, indicative of circumstellar
material, and derived $\sim 1$ Myr as the age of the group.

\begin{figure*}
\centering
\includegraphics*[width=0.95\linewidth,bb=20 54 576 546]{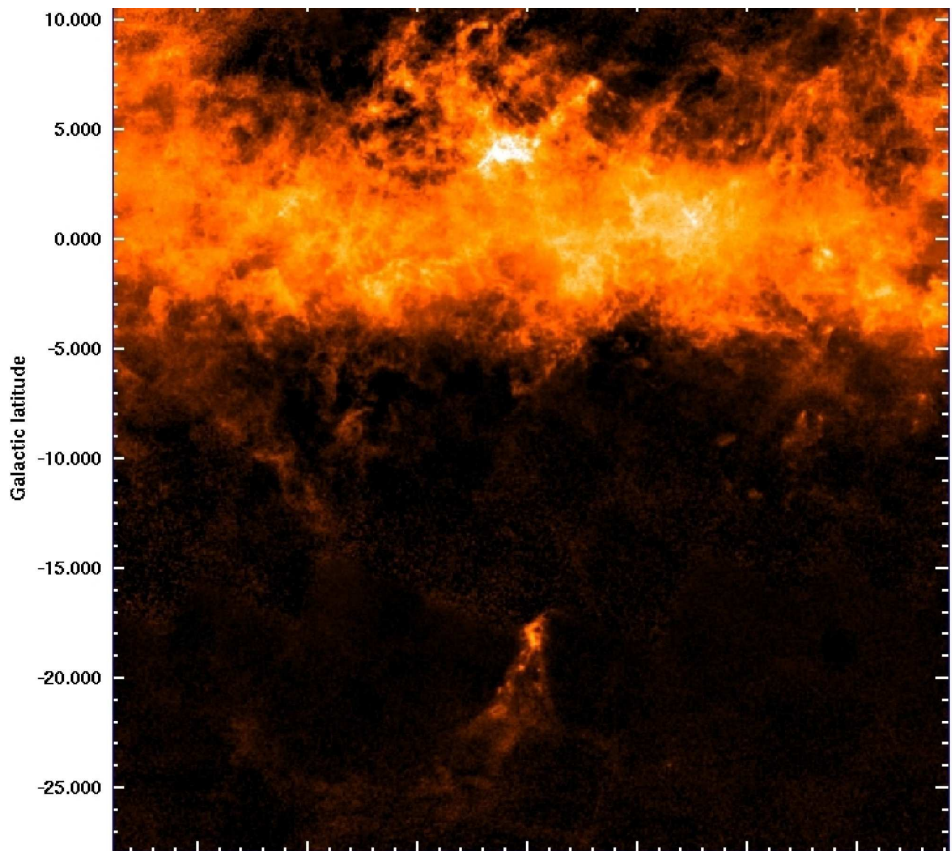}
\caption{The CrA cloud in an extinction map from \citet{dob05}. Note the location relative to the Galactic plane at $b\approx-18^\circ$.}
\label{dobashifig}
\end{figure*}

The Corona Australis molecular cloud complex is today known
as one of the nearest regions with ongoing and/or recent
intermediate- and low-mass star formation.
The dark cloud near R CrA
is the densest cloud core in the region with extinction of up to $A_{V} \sim 45$ mag
\citep{dau87,wil92}.
This cloud is also called {\em condensation A} \citep{ros78}, {\em FS~445 -- FS~447} \citep{fes84}, DoH {\em 2213} \citep[][see Fig.~\ref{dobashifig}]{dob05}, and sometimes also just {\em CrA} or {\em R CrA cloud};
we use {\em CrA} for {\em Corona Australis} as the name for the star-forming region.

\begin{figure*}
\includegraphics*[width=\linewidth,bb=19 23 575 266]{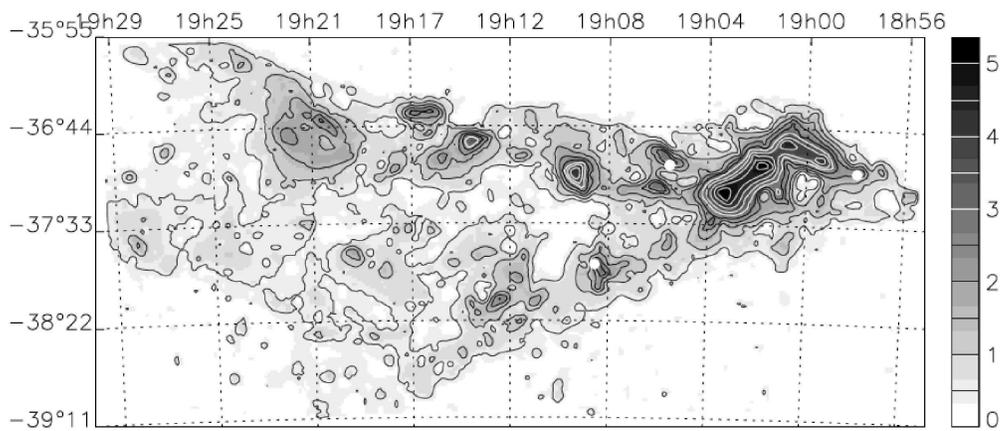}
\caption{Extinction map of CrA from $B$-band star counts (from \citealp{cam99}).}
\label{cambresy_fig6}
\end{figure*}

\subsection{The Galactic Environment}

The CrA dark cloud is located $\sim 18^{\circ}$ below the Galactic plane (see Fig.~\ref{dobashifig}).
According to \citet[see also Fig. 1.10 in \citealp{poe97}]{ola82}, the CrA
cloud is not part of the Gould Belt nor of the Lindblad ring \citep{poe97},
but is located within the massive HI shell called {\em Loop I} \citep{har93}.
The 3D space motion of the CrA T~Tauri stars would be consistent with the
dark cloud being formed originally by a high-velocity cloud impact onto
the Galactic plane, which triggered the star formation in CrA
\citep{neu00}. When a high-velocity cloud impacts onto the gas and dust in the Galactic plane, this
high-velocity cloud may get distroyed, but then forms
a new (lower velocity) cloud.
Alternatively, the star formation in CrA may have been caused by the expansion of the UpperCenLupus (UCL) superbubble: \citet{mam01}
found that the CrA complex is radially moving away from UCL
with 7~km~s$^{-1}$  so that it was located near its center $\sim 14$~Myrs ago.

\citet{har93} resolved cloud {\em A} into five condensations,
the star R~CrA being located in {\em A2}.
\citet{anv96} and \citet{cam99} mapped the cloud using optical star counts (see Fig.~\ref{cambresy_fig6}).
\citet{juv08} study the relationship between scattered near-infrared light and extinction in a quiescent filament of the CrA cloud and compare
the result to extinction measurements.
Several IR surveys revealed a large population of
embedded IR sources \citep{tas84,wil86,wil89,wil92,wil97}
some of which are IR Class~I objects.
Based on their $^{13}$CO map, \citet{tac02} estimated the star formation
efficiency in CrA to be $\sim 4~\%
$.

\subsection{The Distance}

The distance towards the CrA star forming region was estimated
by \citet{gap36} to be $150 \pm 50$~pc and later
by \citet{mar81} to be $\sim 129$~pc (assuming a ratio of total to selective absorption of $R=4.5$, see \citealp{vrb81}).
The \textsl{Hipparcos} satellite attempted to measure the parallax of the
star R CrA and found $122 \pm 68$ mas, i.e. no reliable solution.
\citet{knh98} gave $\sim 170$ pc as distance.
Three likely CrA members have reliable \textsl{Hipparcos} parallax measurements and thus distances; these are HR~7170 ($d$=$85\pm26$~pc), SAO~210888 ($d$=$193\pm39$~pc), and HD~176386 ($d$=$139\pm22$~pc). R~CrA and HR~7169 do not have reliable parallax measurements. The weighted mean of these three distances is $128\pm26$~pc.
\citet{cas98} determined the distance towards the
eclipsing double-lined spectroscopic binary TY CrA
to be $129 \pm 11$~pc from their orbit solution.
The orbit solution of TY~CrA provides the currently best estimate for
the distance to the star-forming region, and we therefore suggest to
use a distance of 130~pc for CrA, see also \citet{dez99}.

\section{Optical Observations and Stellar Members}

Table~\ref{bigtab} gives a list of all known optically visible young members in CrA,
with their names, PMS types (pre-main sequence Herbig Ae/Be or T Tauri stars), spectral types,
H$\alpha$ and lithium equivalent widths, radial
velocity and remarks on binarity (see also \citealp{che97} and \citealp{neu00} for earlier listings). The early work is reviewed by \citet{gra91}.

\subsection{Early Surveys}

Only a few low-mass PMS stars, especially T~Tauri stars (TTSs),
associated with the CrA dark cloud have been found in early H$\alpha$ and IR surveys
\citep[][cf. first \textsl{Spitzer} results in \citealp{fpr06}]{kna73,gla75,mar81,wil86,wil89,wil92,wil97}, all being classical TTSs (cTTSs)
with IR excess and strong H$\alpha$ emission (see Table~\ref{bigtab}).
\citet{pat98} obtained optical photometry and spectroscopy of some more
sources of, at that time, unknown nature around R CrA, previously found by
\citet{kna73}, \citet{gla75}, and \citet{mar81},
and of some X-ray sources found in a pointed ROSAT observation.
He classified some of them as new association members
due to H$\alpha$ emission.
In an optical spectroscopic study of five bright variable stars in the region, \citet{gra92} found rapid changes, interpreted as due to variable accretion, towards R~CrA and the double T~Tauri star S~CrA (see also \citealp{app86,wal05,car07} for this source).

\subsection{The Early-type Members of CrA (Herbig Ae/Be Stars)}

CrA was found to be a young star-forming region mostly
because of the variable stars R and T CrA, which have a
short main sequence contraction time \citep{her60}.
They have spectral types A5e II and F0e, respectively, and
are relatively massive, bright and, hence, were known first.
There are several more such stars in CrA,
most notably TY CrA (B8-9e), but HR 7169, HR 7170, SAO 210888,
and HD 176386 may also be CrA members (see \citealp{neu00}).
Very high resolution ($R \simeq 50000$) H$\alpha$ line profiles
of several Herbig Ae/Be stars (HAeBes) can be found in \citet{rei96}.

In particular the HAeBe star TY CrA is very important for the
CrA cloud (distance estimate) and young stars in general
(mass estimate), because it is an eclipsing double-lined binary:
\citet{gap32}, \citet{kar79}, and \citet{kar81} found TY CrA to show eclipses.
\citet{pro81} noticed a third object in the system,
a visual companion to the (visually unresolved) eclipsing binary.
\citet{hil92}, \citet{wil92}, \citet{bib92},
and \citet{fri96} noticed the IRAS excess emission
in the far IR and concluded that there is a massive accretion disk around
TY CrA. \citet{cas93} and \citet{lag93} obtained high-resolution
spectra, found TY CrA to be a double-lined spectroscopic binary and determined the orbit.
Eclipsing double-lined binaries (eSB2) are very important,
because they can be used to derive the masses of both component stars
directly and dynamically just from Kepler's third law without further assumptions. This derivation was performed by
both \citet{cor94} and \citet{cas95}.
Such stars with dynamical masses are most important for
testing and calibrating pre-main sequence models and our
understanding of star formation is general. The lower-mass component
in TY CrA (B) was the first TTS with a dynamic mass estimate,
while the primary (A) in this system has intermediate mass.
After some improvements (e.g. \citealp{cor96,beu97}),
the current values are $1.64 \pm 0.01$ (B) and $3.16 \pm 0.02~M_{\odot}$ (A),
respectively, for the two stars in the eSB2 TY CrA \citep{cas98}.
\citet{cha03} found a fourth object in the system
using adaptive optics (VLT/NaCo).
The non-eclipse photometric variability was studied by \citet{vaz98}.
The neighboring HAeBe HD~176386 was found to show signs of accretion in observations carried out with the \textsl{International Ultraviolet Explorer} \citep{gra93}.

TY CrA was also among the first intermediate-mass stars towards which hard
X-ray emission was detected, namely by the Japanese satellite
ASCA \citep{koy96}; see \citet{ste06} for a
recent discussion of X-ray emission of HAeBes.

R and T CrA were among the first intermediate-mass young stars towards which
jets and disks were discovered \citep{war85}.
They were also among the first stars where binarity was discovered
by the spectro-astrometric method \citep[][tentative for R CrA]{bai98,tak03}.
R CrA was of course the first variable star noticed in CrA (see Section~\ref{sec_disc}).

In-between the stars R and T~CrA, there is the reflection
nebula NGC 6729 (e.g. \citealp{gra87}), and the stars TY CrA and HD 176386 illuminate the nebula NGC 6726/6727. Studying lines of CN, CH, and CH$^+$ in the direction of TY~CrA, \citet{car89} find that the dust in the region which is attenuating the UV emission is highly processed. \citet{che93} noted strong extended emission at 3.3~$\mu$m in NGC 6726, possibly due to polycyclic aromatic hydrocarbons (PAHs).

\subsection{Mass Function and Optical Follow-up of X-ray Sources}

Since there are several early-type intermediate-mass stars
in CrA, see previous section, there should be many more than
the two dozen TTSs known until the early 1980s, if its initial mass
function (IMF) is consistent with the Miller-Scalo IMF \citep{mis79}.

With optical follow-up observations of previously unclassified X-ray sources
detected with the {\em Einstein} observatory, \citet{wal86} and
\citet{wal97} found eleven new TTS members, namely CrAPMS 1 to 9,
some of them multiple (see Table~\ref{bigtab}).
With only one exception (the cTTS CrAPMS 7), all
of them are weak-emission line TTSs (wTTSs).

From the spatial incompleteness of the {\em Einstein} observations and the
X-ray variability of TTSs, \citet{wal97} concluded that there
should be $\sim 70$ TTSs in CrA.

From their IR survey, \citet{wil97} estimated the number of the
low-mass members to be 22 to 40 for an association age of $\sim 3$ Myrs.
If star formation has been ongoing in CrA for more than $\sim 3$ Myrs,
there should be even more PMS stars. Such older PMS
stars
should in part be found around
the CrA dark cloud, because they had enough time to disperse.

With optical follow-up observations of previously unidentified ROSAT
X-ray\linebreak sources, \citet{neu00} could
identify 19 new PMS stars in and around the CrA cloud (see Table~\ref{bigtab}),
identified as such by low- to intermediate resolution spectroscopy
showing late spectral types, H$\alpha$ emission (or emission filling-in
the absorption), and lithium 6708~\AA~absorption, a youth indicator;
stars with more lithium than zero-age main-sequence stars
such as the Pleiades of the same spectral type are younger, hence
PMS. \citet{bow05} continue this work with optical
follow-up observations of the deep ROSAT PSPC and HRI
pointings, where they may find additional fainter
and later-type members.
See \citet{neu97} for an early review on PMS
identification of ROSAT sources.

For their CrA targets, \citet{jam06} found an age of $9 \pm 4$~Myrs
from high-resolution spectra, placing the stars into the HR diagram
(see also \citealp{nis05} who determine ages of accreting protostars in CrA,
finding that IRS~1 is the youngest at an age of $\sim 10^5$ yrs).
They determined metallicities for three TTSs, found to be slightly sub-solar.


Among the ROSAT X-ray sources in and around the CrA cloud, one particularly
bright but unidentified source was found (RXJ1856.5-2754). After deep
optical follow-up observations, it was identified as a 26th mag
optically visible isolated neutron star \citep{wal96,net97,wam97}. {\em Hubble Space Telescope} parallax observations indicate a distance between 100 and 200~pc \citep{wal02,pos07,kap07}. Due to its space motion, it is considered to come from the Upper Scorpius or Upper Scorpius Lupus OB associations about 0.4~Myr ago \citep{wal01}.

\subsection{3D Space Motion}

\citet{wal97} and \citet{neu00} established the
typical radial velocity of kinematic members of the CrA association
to be $-2$ to 0~km~s$^{-1}$ (heliocentric) in the cloud core
(confirmed by \citealp{jam06}) and $-5$ to 0~km~s$^{-1}$ for TTSs around the cloud. \citet{tei00} list proper motions for six PMS stars in the region.

The mean proper motion of the optical members with known proper motion is
$(\mu _{\alpha} \cdot \cos \delta, \mu _{\delta}) \simeq (5.5, -27.0)$
mas/yr \citep{neu00}, which is mainly solar reflex motion.

\subsection{Rotation}

While for most of the new TTSs discovered by ROSAT (see \citealp{neu00})
as well as for the previously known TTSs in CrA, projected equatorial rotational
velocities $v \cdot \sin i$ are known, only very few rotational periods
were measured in CrA so far.

\citet{cov92} monitored CrAPMS 1, 2, and 3 (their stars Wa/CrA 1, 2, 3)
as well as S CrA at the ESO La Silla 50cm telescope in Chile
and found a period in CrAPMS 2 of 2.9 days, which is typical for wTTSs.

\citet{shev95} found tentative periods for CrAPMS 1 and 2,
namely 2.24 and 2.79 days, respectively, with less significant
false alarm periods of 1.79 and 1.55 days, respectively.
Then, \citet[]{foe05} monitored the field around the \textsl{Coronet} Cluster
with several TTSs in the field in B- and H$\beta$ filters; he
found a new period of $\sim 3.75$ days with the Fourier
transform technique for CrAPMS 3 with a less significant (false alarm)
peak at 1.36 days.

These few known periods are typical for wTTSs (up to a few days)
and cTTSs (more than a few days) found in other star-forming regions.

\begin{table}
\footnotesize
\caption{All known, young, optically visible members of CrA}
\label{bigtab}
\begin{tabular*}{\textwidth}{@{\extracolsep{-1.8mm}}lllllll}
\noalign{\smallskip}
\tableline
\noalign{\smallskip}
Designation, other names$^{(1)}$ & $^{(2)}$ & Spec.        & $W_{\lambda}(H_\alpha )$ & $W_{\lambda}$(Li) & RV       & multiplicity \\
                                 &         & type         & [\AA ]                   & [\AA ]            & \multicolumn{2}{l}{[km~s$^{-1}$]}  \\
\noalign{\smallskip}
\tableline
\noalign{\smallskip}
S CrA & c & K6 	  & $-90.0$ & 0.39$^{{a}}$ & 0	  & $1.34''$ bin$^{{b},{c},{d}}$ \\
TY CrA  & H & B9e &    & yes	  &	  & quad.$^{{e},{f},{g},{h},{i}}$ \\
R CrA & H & A5e II       &	 &	       & -36$^{(3)}$ & bin$^{{j},{k}}$ \\
DG CrA & c & K0$^{{a}}$	 & $-77.9^{{a}}$ & 0.57$^{{a}}$ & & \\
T CrA & H & F0e 	 & & & & $\ge 0.14''$ bin$^{{c}}$ \\
VV CrA & c & K1$^{{a}}$	 & $-72.0^{{a}}$ & & & $2.08''$ bin$^{{l},{d}}$ \\
CrAPMS 7   & c & M1		 & $-33.5$ & 0.36 & & \\
V709 CrA   & w & K1  	 & $0.3$ & 0.39 & -1.0 & \\
HBC 677 & c$^{{a}}$ & K8-M2$^{{m}}$ & $-46.0^{{a}}$ & $0.47^{{a}}$ & & \\ 
V702 CrA & w & G5  	 & $-0.7$ & 0.28 & -1.2 & \\
CrAPMS 3  & w & K2  	 & $-0.9$     & 0.41 & -1.2 & $4.45''$ bin$^{{l},{d}}$ \\
CrAPMS 3/c	  	& w$^{{a}}$ & M4$^{{a}}$	 & $-6.8^{{a}}$ & $0.36^{{a}}$ & & $4.45''$ bin$^{{l},{d}}$ \\
HBC 680       & w$^{{a}}$ & M0-3$^{{a},{m}}$     & $-4.6^{{a}}$ & 0.64$^{{a}}$ & & $0.22''$ bin$^{{d}}$ \\ 
{[GP75]} R CrA e2 & T & M3-5$^{{n}}$   & em.$^{{n}}$ & & & \\
{[GP75]} R CrA f2 & w & K4$^{{n}}$	 & $0.1^{{a}}$ & $0.44^{{a}}$ & & $^{(4)}$ \\
{[GP75]} R CrA n & c$^{{n}}$ &	 & em.$^{{n}}$ & & & \\
{[MR81]} HA 12  & T & M3-5$^{{n}}$   & em.$^{{n}}$ & & & \\
{[MR81]} HA 13  & T & M3-5$^{{n}}$   & em.$^{{n}}$ & & & \\
{[MR81]} HA 15  & T & M3-5$^{{n}}$   & em.$^{{n}}$ & & & $0.22''$ bin$^{{d}}$\\
V721 CrA & w$^{{o}}$ & K$^{{o}}$& $-5.0^{{o}}$ & $0.35^{{o}}$ & & \\
{[MR81]} HA 17  & T & M3-5$^{{n},{p}}$ & em.$^{{n},{p}}$ & & & $1.3''$ bin$^{{d},{p}}$ \\
CrAPMS 4NW & w & M0.5	 & $-1.1$  & 0.45 & -2.2 & $0.36''$ bin$^{{d}}$ \\
CrAPMS 4SE & w & G5  	 & $1.0$   & 0.36 & -2.0 & \\
CrAPMS 5 & w & K5  	 & $-0.8$  & 0.44 & -0.8 & \\
CrAPMS 6NE & w & M3	 & $-5.9$  & 0.70 & & $2.37''$ bin$^{{q},{d}}$ \\
CrAPMS 6SW & w & M3.5	    & $-9.8$  & 0.44 & & $2.37''$ bin$^{{q},{d}}$ \\
CrAPMS 8 & w & M3  	 & $-3.9$  & 0.57 & & $0.13''$ bin$^{{d}}$ \\
CrAPMS 9 & w & M2  	 & $-9.2$  & 0.5  & & \\
RXJ1855.1-3754 & w$^{{r}}$ &K3$^{{r}}$& $1.6^{{r}}$ & $0.38^{{r}}$ & & \\
RXJ1857.7-3719 & T & M3-5$^{{n}}$   & em.$^{{n}}$ & & & \\
RXJ1858.9-3640 & T & M3-5$^{{n}}$   & em.$^{{n}}$ & & & \\
RXJ1859.7-3655 & T & M3$^{{n}}$	 & em.$^{{n}}$ & & & \\
RXJ1836.6-3451 & w & M0 	  & $ -2.4 $ & 0.54 & $-1.6$& \\
RXJ1839.0-3726 & w & K1 	  & $ +0.1 $ & 0.34 & $-4.8$& \\
RXJ1840.8-3547 & w & M4 	  & $ -6.9 $ & 0.36 &	    & \\
RXJ1841.8-3525 & w & G7 	  & $ +1.0 $ & 0.25 & $-3.1$& \\
RXJ1842.9-3532 & c & K2 	  & $-30.7 $ & 0.38 & $-1.2$& \\
RXJ1844.3-3541 & w & K5 	  & $ -0.4 $ & 0.41 & $-4.9$& $0.227''$ bin$^{{d}}$ \\
RXJ1844.5-3723 & w & M0  	 & $ -5.3 $ & 0.52 &	   & \\
RXJ1845.5-3750 & w & G8  	 & $ +0.8 $ & 0.24 & $-1.3$ & $0.9''$ bin$^{{d}}$ \\
{[NWC2000]} RXJ1846.7-3636~NE 	      & w & K6  	 & $ -0.6 $ & 0.47 &$-2.5$ & $0.17''$ bin$^{{d}}$ \\
{[NWC2000]} RXJ1846.7-3636~SW 	      & w & K7  	 & $ -0.2 $ & 0.43 &$-2.6$ & $^{(6)}$ \\
RXJ1852.3-3700 & c & K3  	 & $-33.8 $ & 0.51 & $ 0.0$ & \\
RXJ1853.1-3609 & w & K2  	 & $ +1.4 $ & 0.39 & $-3.2$$^{s}$ & $0.516''$ bin$^{{d}}$ \\
RXJ1856.6-3545    	      & w & M2  	 & $ -1.6 $ & 0.25 &	   & $3.65''$ bin$^{{d}}$ \\
{[NWC2000]} RXJ1857.5-3732~E  	      & w & M5  	 & $ -3.4 $ & 0.23 & $-3.1$& $2.86''$ bin$^{{a},{d}}$ \\
{[NWC2000]} RXJ1857.5-3732~W  	      & w & M6  	 & $ -4.7 $ & 0.42 & $-2.9$& $2.86''$ bin$^{{a},{d}}$ \\
RXJ1901.4-3422 & w & F7  	 & $ +2.2 $ & 0.09 & $-3.3$& $^{(5)}$ \\
RXJ1901.6-3644 & w & M0  	 & $ -2.3 $ & 0.54 &	   & $0.935''$ bin$^{{d}}$ \\
RXJ1917.4-3756 & w & K2  	 & $ -0.9 $ & 0.48 & $-0.3$$^{{s}}$ & $0.14''$ bin$^{{d}}$ \\
\hline
\end{tabular*}
\end{table}
\begin{table}
\footnotesize
{Table~\ref{bigtab}, cntd.}

\begin{tabular*}{\textwidth}{@{\extracolsep{-1.8mm}}lllllll}
\tableline
Designation, other names$^{(1)}$ & $^{(2)}$ & Spec.        & $W_{\lambda}(H_\alpha )$ & $W_{\lambda}$(Li) & RV       & multiplicity \\
                                 &         & type         & [\AA ]                   & [\AA ]            & \multicolumn{2}{l}{[km~s$^{-1}$]}  \\
\noalign{\smallskip}
\tableline
\noalign{\smallskip}
LS-RCrA 1	 	     & c & M6.5$^{{t}}$	& large$^{{t},{u},{v}}$  & $0.74^{{u}}$   & $2 \pm 3^{{u}}$ &   \\
LS-RCrA 2	 	     & w & M6$^{{t}}$	& $-1.7 ^{{l}}$  &     &       & \\
DENIS-P~J185950.9-370632 &b$^{{w}}$&M8& $-18$ & 0.41 & & $0.06''$ bin$^{{w}}$ \\
{[LEM2005b]} CrA 432 	 	     & b$^{{x}}$ & M7$^{x}$     & large$^{{x}}$ &&& \\
{[LEM2005b]} CrA 444 	 	     & b?$^{{x}}$ & M8.5$^{{x}}$		 &	       &&& $3''$ bin?$^{{x}}$\\
\hline
\end{tabular*}

\smallskip

Comments: \\
The 2nd column gives the PMS type (see below).
H$\alpha$ equivalent widths are negative when in emission. RV stands for radial velocities.
The last column gives information on the multiplicity with the separation in arcsec for
binaries with reference(s) to first discovery of component(s) and most recent and/or precise separation value(s),
or otherwise as noted.
Data for stars with CrAPMS designations are from \citet{wal97},
for most RXJ designations from \citet{neu00}, for other stars with HBC
number from the Herbig-Bell catalog (HBC, \citealp{her88}), unless otherwise noted.
MR81 H$\alpha$ 10 and Kn anon 2 were previously listed as PMS stars or
candidates in CrA, but found to be non-TTSs in \citet{neu00}.
Additional low-mass optically visible members may be Patten R7b (K4), {[GP75]} i2 = VSS VIII-13 (K2),
and {[GP75]} wa = Patten R10 (M2), all according to Meyer \& Wilking (2003).
Additional intermediate-mass late B-type Herbig Ae/Be star members
may be SAO 210888, HR 7169, HR 7170, and HD 176386 \citep{neu00}.

\smallskip

Notes: \\
$^{(1)}$ GP or {[GP75]} for Glass \& Penston 1975, MR or {[MR81]} for Marraco \& Rydgren 1981. \\
$^{(2)}$ PMS type: c for cTTS, w for wTTS, T for TTS (not known
whether cTTS or wTTS), H for Herbig Ae/Be stars, and b for brown dwarf.\\
$^{(3)}$ The radial velocity given by \citet{wil53} of --36~km/s
(error given as  $<5$~km/s, quality code \textsl{o}, based on only two plates) would be
inconsistent with CrA membership, hence seems dubious, or may indicate binarity.\\
$^{(4})$ {[GP75]} f2 is identical to RXJ1901.1-3648 listed twice in \citet{neu00}.\\
$^{(5)}$ RXJ1901.4-3422 at $65 \pm 5$ pc seems to be a young PMS foreground star (see \citealp{neu00}). \\
$^{(6)}$ RXJ1846.7-3636 SW and NE have a separation of 7.884 arc sec,
so that that may be a bound binary, but are not counted as binary here because of their large separation,
RXJ1846.7-3636 NE is itself a close binary with 0.17 arc sec separation. \\

\smallskip

References:
(${a}$) \citet{neu00},
(${b}$) \citet{joy44},
(${c}$) \citet{bai98},
(${d}$) \citet{koe08},
(${e}$) \citet{gap36},
(${f}$) \citet{pro81},
(${g}$) \citet{cas98},
(${h}$) \citet{cha03},
(${i}$) visual companion has spectral type K0 according to \citet{mw03},
(${j}$) not an astrometric binary according to \citet{bai98},
(${k}$) spectro-astrometric binary \citet{tak03},
(${l}$) \citet{rei93},
(${m}$) \citet{mw03},
(${n}$) \citet{pat98},
(${o}$) \citet{grh02},
(${p}$) Neuh\"auser et al. \textsl{in prep.},
(${q}$) \citet{wal97},
(${r}$) \citet{net97},
(${s}$) \citet{jam06},
(${t}$) \citet{fer01},
(${u}$) \citet{bar04},
(${v}$) \citet{moh05},
(${w}$) binary brown dwarf with primary having strong (i.e. cTTS-like) H$\alpha$ emission \citet{bou04},
(${x}$) \citet{lop05}.
\end{table}

\subsection{Multiplicity (and a New Edge-on Disk)}

Only few stellar multiples are published among the CrA members so far,
namely the binaries S~CrA, R CrA (dubious case, \citealp{bai98}, \citealp{tak03}),
T CrA, VV CrA \citep[which has a reddened companion, see][]{rei93},
CrAPMS 3, CrAPMS 6, {[MR81]} H$\alpha$ 11, RXJ1846.7-3636,
RXJ1857.5-3732 and the quadruple TY CrA -- few compared to the total number of 46 member systems (excluding brown dwarfs).

\citet{koe08} performed a multiplicity
survey with the MPE speckle camera SHARP and the AO camera ADONIS
by observing most of the members listed in Table~\ref{bigtab};
they discovered 13 new binaries.
Among the 7 Herbig Ae/Be stars known in CrA, there are now
six multiples (quadruple TY CrA, triple HR 7170,
and binaries HR 7169, R CrA, T CrA and HD 176386; HR 7170 and HR 7169 may together form a quintuple).
Among the 43 TTS, there are 21 binaries, but no higher order multiples
(RXJ1846.7-3636 NE A-B and SW are either a binary and a single TTS, or a triple system).
Given 23 multiples among 50 systems, we have a percentage of $46 \pm 10~\%$
of multiplicity. Counting all binaries within triples and quadrules, 26
binaries in 50 systems give a percentage of $52 \pm 10~\%$ binaries.
This is a relatively high percentage of multiples,
considering the fact that neither all members are surveyed
nor that all separations are probed.
Hence, it seems likely that the multiplicity in CrA
is normal, i.e., as high as in similar star forming regions.

\begin{figure*}
\centering
\includegraphics[width=8cm]{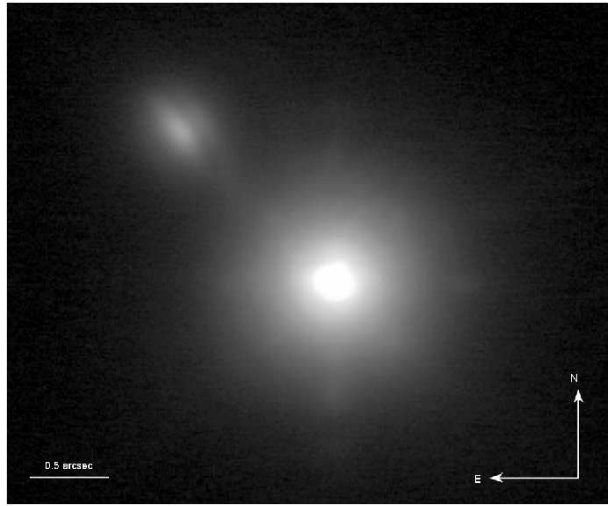}
\caption{VLT NaCo AO \textsl{K}-band image of the binary {[MR81]} H$\alpha$ 17
with one component surrounded by an (almost) edge-on disk seen in the upper left
located 1.4 arcsec north-east of the apparently brighter component. Note the faint dust
lane of the disk (Neuh\"auser et al., \textsl{in prep.}).}
\label{mugifig}
\end{figure*}

During the multiplicity survey by \citet{koe08},
the mid-M type member {[MR81]} H$\alpha$ 17 was found to be binary
with the fainter companion being so faint that it could have
been a sub-stellar companion, so that follow-up observations
have been performed already with the VLT, both spectroscopy
of both components as well as imaging with the Adaptive Optics
camera NaCo. The image with very high dynamic range and spatial
resolution revealed an (almost) edge-on disk around the apparently
fainter object, which may be the earlier-type more massive
object, appearing fainter because of the disk extinction,
see Fig.~\ref{mugifig} (Neuh\"auser et al., \textsl{in prep.}).

Most recently, R CrA was observed in the mid-infrared with the VLT
Interferometer (VLTI) by \citet{zin05}, who resolved circumstellar material
on a 6--10 AU scale, a ring with $\sim 45^{\circ}$ inclination.

\subsection{Herbig-Haro Objects}

Starting in the 1970s, Herbig-Haro objects were discovered in the CrA star-forming region. Seven such objects were discovered until the early 1990s, namely HH~82, 96--101, and 104 \citep{str74,coh84,sjs84,har87,rei88,grx93}. In a comprehensive survey carried out with the Wide-Field Imager at the ESO/MPG 2.2m telescope, \citet{wan04} discovered twelve new HH objects, mainly surrounding R~CrA and VV~CrA, while confirming all earlier detections.

\section{Infrared Surveys}

Following earlier observations of the brighter stars R~CrA, T~CrA, and
TY~CrA, \citet{kna73} presented infrared data of eleven sources in the
region, finding many cases of infrared excess emission. \citet{vrb76}
found 18 sources with a 3$\sigma$ limit of 10 mag in the $K$ band. Two
extended far-infrared sources, around R~CrA and around TY~CrA, were
found by \citet{cru84}. With a limiting detectable magnitude of
$K=14$, \citet{tas84} discovered a cluster of embedded sources in the
vicinity of R~CrA, naming it the \textsl{Coronet} cluster. They listed
sources IRS~1--15. After far-infrared studies of the region
\citep{wil85}, \citet{wil86} were able to distinguish association
members from field stars using mid-IR (MIR) photometry. They found 13
members in many different evolutionary stages. While \citet{wil92}
identified 24 Young Stellar Objects (YSOs) in the region using IRAS
and better MIR data, \citet{wil97} listed 22 to 40 association members
among 692 sources detected in deep $JHK$ imaging (down to a
completeness limit of $K=16.5$). \citet{che93} examined the 3.8~$\mu$m
water-ice absorption feature in 11 young CrA objects and eight
background stars, concluding that ice condensation onto grains has
occurred throughout the cloud (see also \citealp{tan94} who study
water-ice and additionally CO ice absorption at 4.7~$\mu$m in
CrA). Using ISOCAM onboard the \textsl{Infrared Space Observatory} for
a survey of the R~CrA region, \citet{olo99} found 24 YSOs due to their
MIR excess emission. Out of these sources, only seven were identified
as YSOs before. The central region of the cloud had to be taken out of
the survey due to saturation problems. Selecting five YSOs from the
\textsl{Coronet} cluster including the three most prominent Class~I
protostars, \citet{nis05} performed detailed near-IR (NIR)
spectroscopy and find very different accretion properties, even for
sources with similar ages. \citet{haas08} present a near-infrared
survey of the entire CrA cloud complex, identifying new deeply
embedded members in the Coronet region.

\subsection{The \textsl{Coronet} Cluster}

\begin{figure*}
\includegraphics*[width=\linewidth, bb=20 53 576 398]{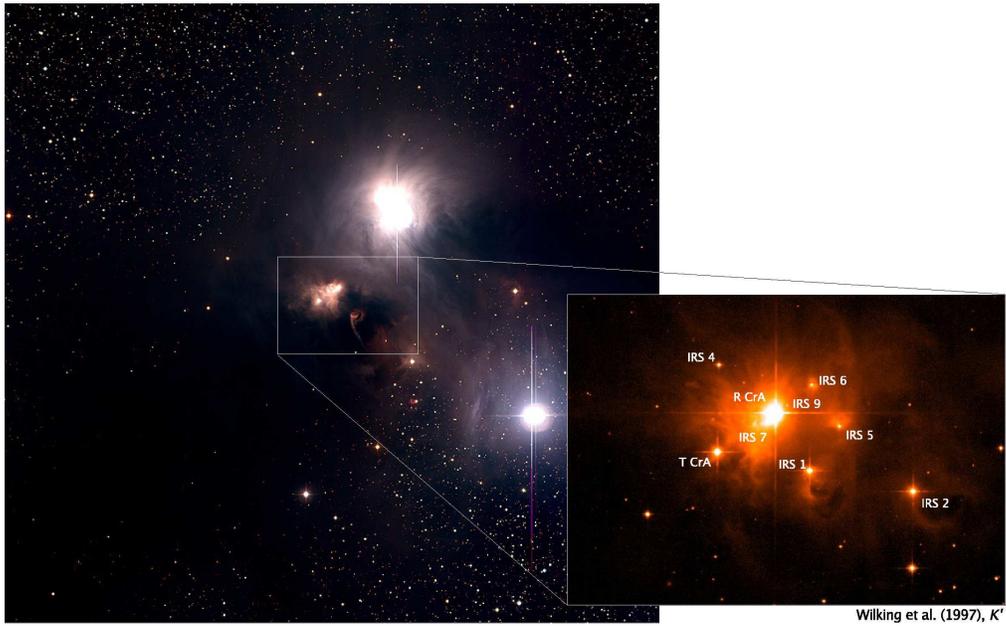}
\centering
\includegraphics[width=10cm, bb= 0 0 436 460]{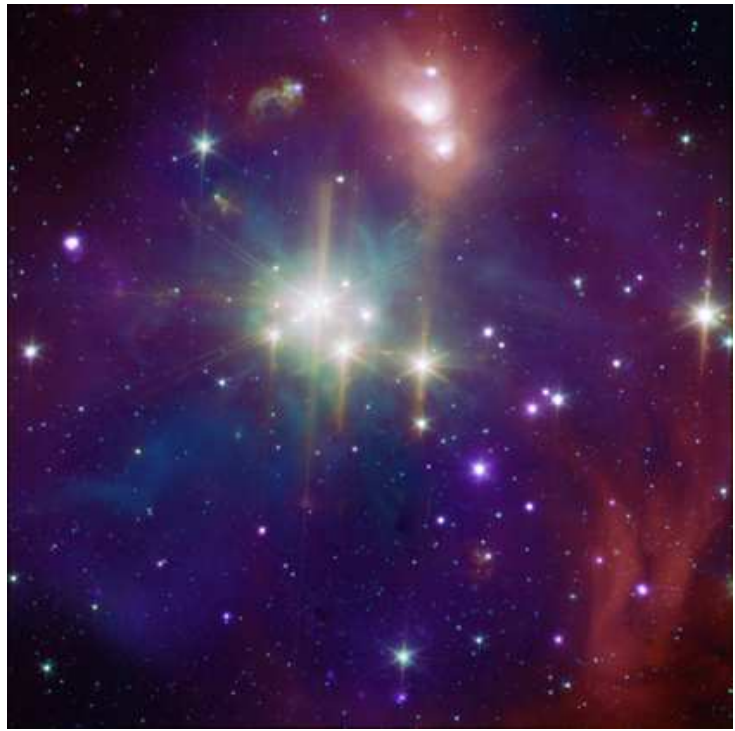}
\caption{Top: An optical image of the CrA region (ESO Press Photo 25a/00). The inset shows the \textsl{Coronet} cluster in $K'$ data from \citet{wil97}. Bottom: This composite image shows the \textsl{Coronet} in X-rays as observed by \textsl{Chandra} (purple) and infrared emission as observed by \textsl{Spitzer} (cyan: 4.5~$\mu$m, green: 8.0~$\mu$m, and orange: 24~$\mu$m). The \textsl{Spitzer} image shows young stars plus diffuse emission from dust. The image is 16.8 arcmin across. Credit: X-ray: NASA/CXC/CfA/J.Forbrich et al.; Infrared: NASA/SSC/CfA/IRAC GTO Team}
\label{corocolor}
\end{figure*}

\begin{figure*}
\centering
\includegraphics[angle=-90,width=8.5cm]{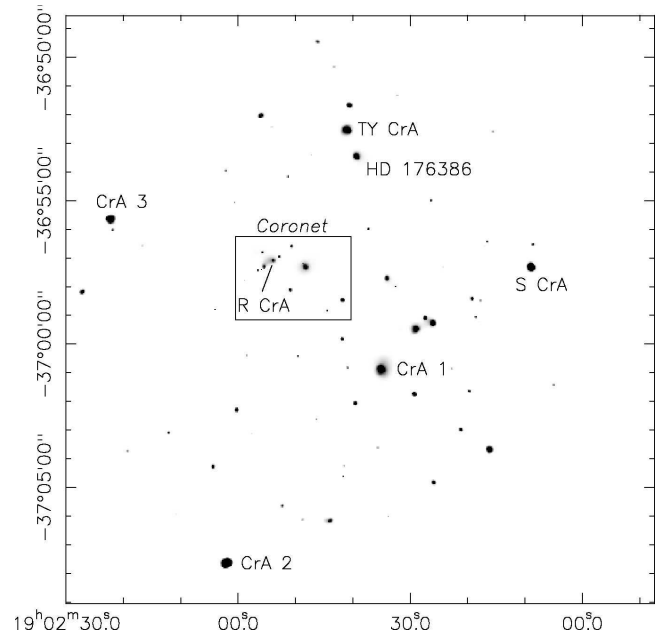}
\caption{\textsl{Chandra}-ACIS X-ray image of the CrA region. This merged dataset with a total exposure time of 156~ksec is one of the most sensitive X-ray images ever obtained of a star-forming region \citep[][and references therein]{fpr06}. Note that there is no data in the corners of this image. Coordinates are J2000.}
\label{plotacis19}
\end{figure*}

The \textsl{Coronet} cluster has been an interesting target since its discovery \citep{tas84}. It contains a very compact
grouping of protostars in early evolutionary stages (see Fig.~\ref{corocolor}), many of them detected at X-ray and radio wavelengths. A list
of the X-ray--detected YSOs in the \textsl{Coronet} cluster can be found in Table~\ref{ysotable} (see also Fig.~\ref{plotacis19} and Section~\ref{sec_xray}). A few of the infrared
Class~I objects were the first such sources detected in X-rays by ASCA and ROSAT observations \citep{koy96,nep97}.
\citet*{for05} analyzed radio and X-ray data of most sources in the \textsl{Coronet}
cluster, finding many signs of variability. Subsequent simultaneous multi-wavelength
observations were carried out in August 2005 \citep{for06}. \citet{fpr06} discuss membership in the \textsl{Coronet} region based on deep X-ray and mid-infrared imaging data.

\begin{table*}
\caption[]{X-ray-detected protostars in the \textsl{Coronet} cluster}
\label{ysotable}
\smallskip
\begin{tabular}{lllrrrcr}
\tableline
Object         & Class$^{\rm a}$ & Position (J2000)     & $J$  & $H$  & $K$  & cm$^{\rm b}$ & env.$^{\rm c}$   \\
               &       &                      & mag$^{\rm d}$  & mag$^{\rm d}$  & mag$^{\rm d}$  &       & $M_\odot$ \\
\hline
IRS 2           & I     & 19 01 41.6 -36 58 31 & 13.7 &  9.7 &  7.1 & yes   & 0.3	 \\
IRS 5           & I     & 19 01 48.1 -36 57 22 & 15.8 & 13.7 & 10.1 & yes$^{\rm e}$   & 0.3	 \\
IRS 6           &II/III & 19 01 50.5 -36 56 38 & 16.4 & 12.1 & 10.3 & yes   &		 \\
IRS 1           & I     & 19 01 50.7 -36 58 09 & 16.0 & 11.1 &  7.6 & yes   & 0.3	 \\
IRS 9           & I     & 19 01 52.6 -36 57 01 & --   & 13.4$^{\rm f}$ & 11.7$^{\rm f}$ & no    &		 \\
R CrA          & HAe   & 19 01 53.7 -36 57 08 &  6.9 &  5.0 &  2.9 & yes   &		 \\
IRS 7A          & I     & 19 01 55.3 -36 57 22 & --   & 13.5$^{\rm f}$ & 12.2$^{\rm f}$ & yes   & 0.6		 \\
IRS 7B$^{\rm g}$& 0/I    & 19 01 56.4 -36 57 28 & --   & --   & --   & yes   & 0.2		 \\
\hline
\end{tabular}

($^{\rm a}$) 0/I = Class~0/I protostar, HAe = Herbig Ae,
($^{\rm b}$) centimeter radio \citep{bro87,sut96,fcw98,for05,mie08},
($^{\rm c}$) submillimeter envelope mass from \citet{nut05},
($^{\rm d}$) 2MASS photometry except for IRS 9 and IRS 7A,
($^{\rm e}$) only known Class~I protostar with circularly polarized radio emission \citep{fcw98},
($^{\rm f}$) $JHK$ photometry from \citet{wil97}, close to bright R CrA,
($^{\rm g}$) enormous X-ray flare reported by \citet{ham05} \\

\end{table*}

\subsection{Brown Dwarfs in CrA}

\citet{wil97} found five brown dwarf candidates in their deep IR
survey, classified as such by their IR $JHK'$ values, see Table~\ref{bdtable}.
They were also observed in X-rays by ROSAT, but not detected \citep{net97}.

\citet{fer01} identified two objects near the sub-stellar limit,
LS-RCrA 1 and 2 ({\em LS} for {\em La Silla}), with spectral types
M6.5-7 and M6, respectively. They classify LS-RCrA 2 as a low-mass star
and LS-RCrA 1 as a low-mass star or brown dwarf candidate.
The object LS-RCrA 1 displays multiple
permitted and forbidden lines normally attributed to accretion and
outflows superimposed on an unusually faint continuum \citep{fer01}.
Its location in the H-R diagram may be below all other established
CrA members, which could be due either to strong accretion and/or
strong extinction (e.g. an edge-on disk). Based on asymmetric forbidden
line profiles, in which the blue tail of the line does not have a
corresponding red counterpart, \citet{fer05} suggest
that the receding outflowing gas is blocked from view by a disk which would
thus not be edge-on (see also \citealp{fer01} and \citealp{bar04} for a
discussion).

\begin{table*}
\caption[]{Brown dwarfs and brown dwarf candidates in CrA.}
\label{bdtable}
\begin{tabular}{lll} \\ \hline
Object & Spectral & Reference  \\
names  & type     &            \\ \hline

B185815.3$-$370435 & & \citet{wil97}  \\
B185831.1$-$370456 & M8.5 & \citet{wil97} $\&$\\
                   &      & \citet{lop05}  \\
B185839.6$-$365823 & & \citet{wil97}  \\
B185840.4$-$370433 & & \citet{wil97}  \\
B185853.3$-$370328 & & \citet{wil97}  \\
LS-RCrA 1          & M6.5 & \citet{fer01} \\
LS-RCrA 2          & M6   & \citet{fer01} \\
DENIS-P J185950.9-370632~A & M8 & \citet{bou04} \\
DENIS-P J185950.9-370632~B & (*) & \citet{bou04} \\
{[LEM2005b]} CrA 444 & M8.5 & \citet{lop05}  \\
{[LEM2005b]} CrA 432 & M7   & \citet{lop05}  \\
\hline
\end{tabular}

(*) B is 1.1 to 1.5 mag fainter than A in the optical red and, hence, \\
slightly later in spectral type and lower in mass \citep{bou04}.

\end{table*}

\citet{fcw98} argued that {\em Radio Source 5} (RS 5) from \citet{bro87},
not identical to IRS 5, may be a brown dwarf, if located in CrA and
younger than 10 Myrs given $K$=16.4 mag and $H-K$=1.45 mag \citep{wil97},
however, this depends on theoretical models which are uncertain up to
$\sim 10$ Myrs and the object could also be a highly reddened
background giant. The radio data of \citet{for05}, on the
other hand, shows variability, not typical for evolved giants,
but a characteristic of young objects \citep{for05}.
However, recently performed near-infrared spectroscopy indicates
an extragalactic nature of RS~5 (Forbrich et al., \textsl{in prep.}).

\citet{bou04} pre-selected one brown dwarf candidate from the DENIS
IR survey by $IHK$ colors, namely DENIS-P~J185950.9-370632
(or Denis~1859-37 for short), and classified it as a young brown dwarf
member of CrA based on its spectral type \normalsize{(M8)}, H$\alpha$ emission,
lithium absorption, and proper motion consistent with CrA,
as observed by HST, VLT, Keck, and ISO.
Denis~1859-37 was also detected in USNO, GSC, and 2MASS.
HST images with high dynamic range revealed a close companion
at only $0.06''$ separation (8 AU at 130~pc),
detected at two different epochs (2000 and 2002).
Because separation and position angle did not change as much
as expected from the proper motion of Denis~1859-37~A,
\citet{bou04} concluded that they form a common--proper-motion pair.
The total mass is estimated to be 0.02 to 0.03 $M_{\odot}$.

Denis~1859-37~A and B are the first two confirmed brown dwarfs in
CrA. LS-CrA 1 and 2 are to be considered as brown dwarf candidates,
and the \citet{wil97} brown dwarf candidates are still to be confirmed
as such.  \citet{lop05} identified one more BD and one more candidate
BD (M7--8.5) in a wide-field $RIH_\alpha$ survey carried out with the
ESO/MPG 2.2m telescope on La Silla, the candidate possibly being a
3$"$ binary.  In deep X-ray data, \citet{fpr06} clearly detect three
out of eight candidate brown dwarfs, with an additional two detected
tentatively. \citet{sicilia08} present optical (FLAMES/VLT) and IR
(\textsl{Spitzer}-IRS) spectra of a set of low-mass stars in the
\textsl{Coronet} cluster.

\section{Radio Observations}

\begin{figure}
\centering
\includegraphics[width=12cm]{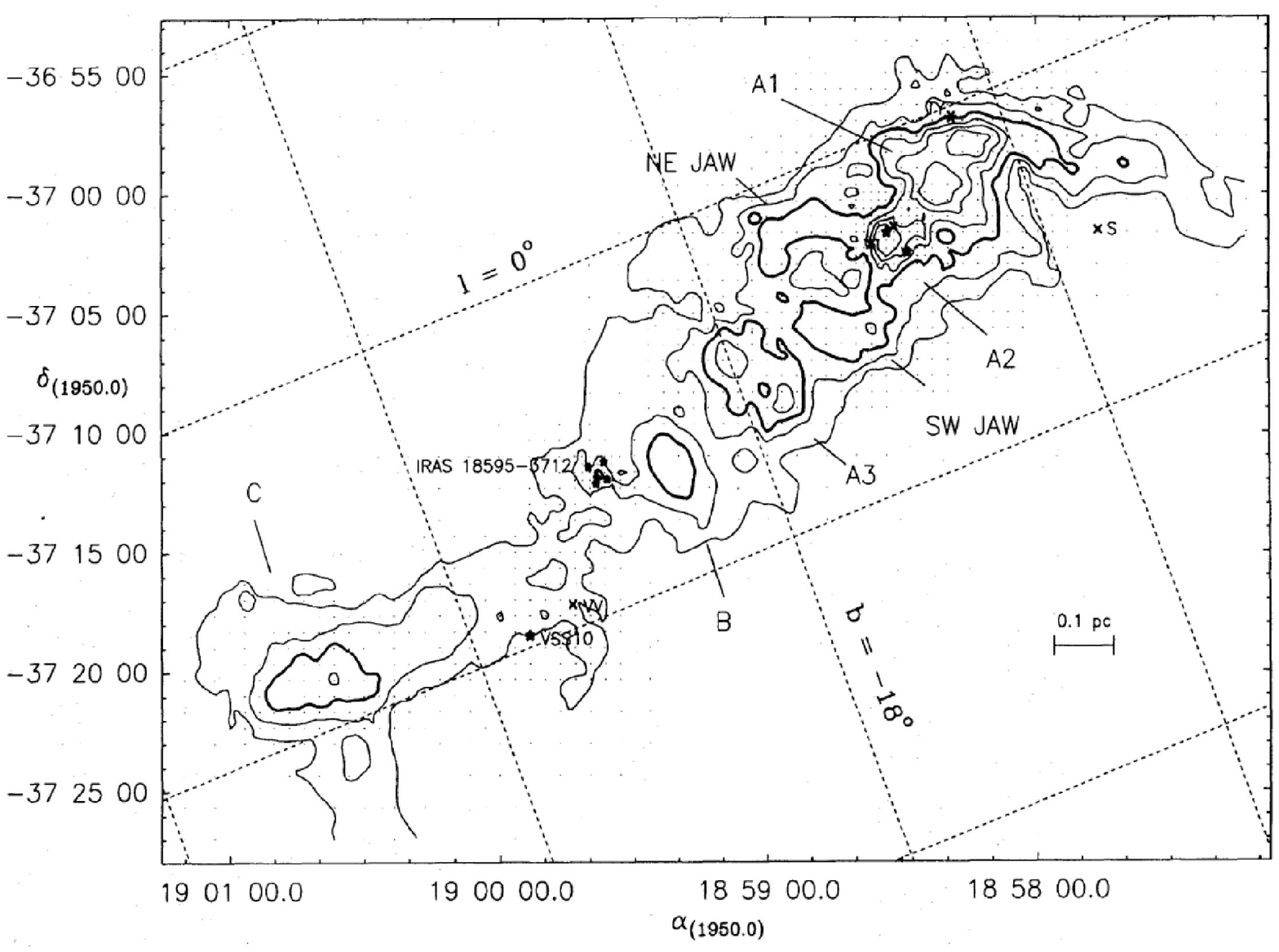}
\includegraphics[width=12.2cm]{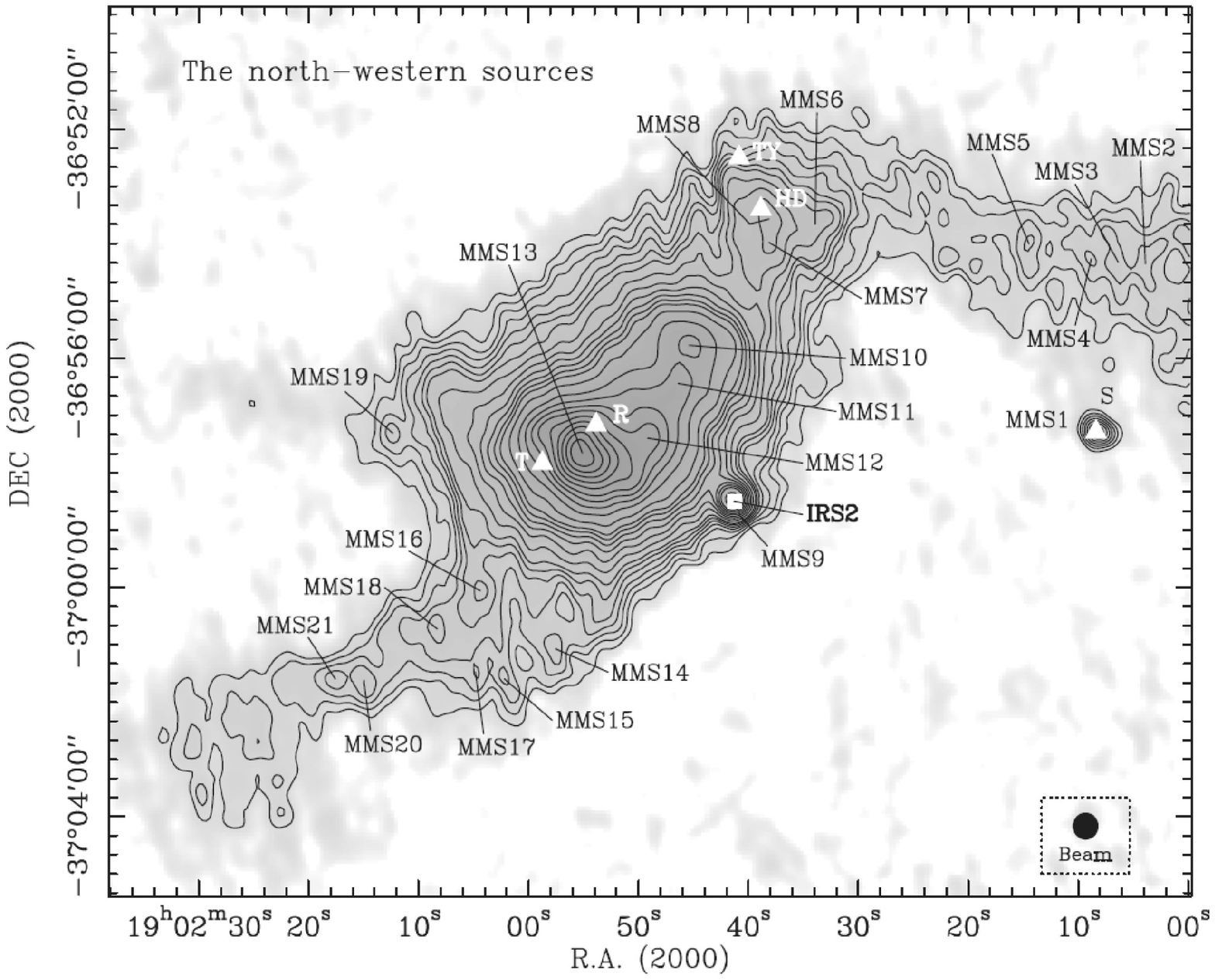}
\caption{Top: Integrated C$^{18}$O(1-0) map from \citet{har93}. Bottom: The north-western part of the cloud in 1.2~mm continuum (from \citealp{chi03}).}
\label{harjuchini}
\end{figure}

The focus of most radio observations of the CrA region until now has
been the vicinity of R~CrA. Already before the discovery of the
embedded sources of the \textsl{Coronet} cluster, molecular line
observations were carried out in the 1970s in order to study star
formation associated with the Herbig Ae/Be stars R CrA and T CrA
(\citealp{lor79} and references therein), and broadened CO line
emission was noticed in the region with a maximum close to the
position of R CrA; \citet{lor80} also detected HCO$^+$ and compared it
to CO. \citet{lev88}, using data of higher resolution, interpreted the
broadened CO emission as a large outflow, noting that it might be due
to both R CrA and the embedded source IRS~7. \citet{har93}, performing
wide-field mapping of the region in the column density and, thus,
mass-tracing molecule C$^{18}$O (see Fig.~\ref{harjuchini}, top
panel), estimated the total gas mass to be around 120~$M_\odot$ while
its densest region around R CrA and IRS~7 contains around half of this
mass (see also \citealp{yon99}).  In a kinematic analysis using
molecular lines, \citet{and97b,and97} found a rotating disk around
IRS~7 with a central mass of 0.8~$M_\odot$, as derived from a fit to
the rotation profile.  \citet{gro04}, using multiple molecular
transitions and 870~$\mu$m continuum data, suggested that the source
of the molecular line and cold dust emission with signs of infall,
rotation, and outflow is not IRS~7 itself, but a Class~0 source in its
vicinity.

In the millimeter continuum, the dust emission at 1.2~mm was studied
by \citet{hen94} who discovered that the emission maximum is not
related to R~CrA or T~CrA but located close to IRS~7 in the
\textsl{Coronet} cluster.  \citet{chi03}, mapping a larger region
around this source, reported several peaks, the brightest one (called
MMS~13) again in the vicinity of IRS~7 in the \textsl{Coronet} cluster
(see Fig.~\ref{harjuchini}, bottom panel). The Class~I protostars
IRS~1, 2, and 5 are associated with millimeter emission as well.  They
concluded that the MMS~13 peak probably is a Class~0 source not
related to any of the sources observed at other wavelengths.
Additionally based on molecular line data, \citet{gro04} reached a
similar conclusion. \citet{nut05} resolved MMS~13 into two
protostellar and one prestellar source at 450~$\mu$m.  The latter
source is at the position of the Class~0 source surmised earlier and
is estimated to have a mass of 6 to 11~$M_\odot$, while the envelopes
of the other protostars all have sub-solar mass dust masses.
\citet{sch06}, observing submillimeter molecular transitions towards
three CrA sources (including IRS~7) with the APEX telescope, found
that the CO fractional abundances are very similar for the sources, in
contrast to H$_2$CO and CH$_3$OH. High angular resolution
submillimeter observations of IRS~7 were presented by \citet{gro07}.

\citet{brz75} detected two sources in the radio continuum (11~cm, NRAO
Green Bank Interferometer), one being close to the position of R CrA.
After the discovery of the \textsl{Coronet} cluster, \citet{bro87}
detected eleven 6~cm radio sources in the area, five of which could be
identified as YSOs. No signs of circular polarization were found.
\citet{sut96} analyzed the multi-band radio variability of these
sources using archival VLA (1985-1987) and Australia Telescope (1992)
data.  They found a major outburst of IRS~5 -- a Class~I protostar --
in 1992, concluding from the spectral index that it was due to a
thermal emission process.  Again, no circular polarization was
found. Also, Radio Source 5 showed signs of variability.
\citet{fcw98} observed the same region at 3.5~cm using the VLA,
discovering circularly polarized emission from IRS~5 while detecting
eight sources.  They concluded that previous detections of circular
polarization failed due to lack of sensitivity and inappropriate
instrumental configuration.  \citet{for05} analyzed short-term radio
variability of these sources using multi-epoch VLA data spread over
nearly four months.  Again, IRS~5 was the most variable source, with
changes seen also in its polarization.  In a deep map, they detected
five additional weak radio sources. In simultaneous multi-wavelength
observations from X-ray to radio wavelengths carried out in August
2005, no clear signs of correlated variability were found
\citep{for06}.  Most recently, \citet{mie08} discuss the Coronet radio
spectral indices while \citet{choi08} analyze archival VLA data of
the Coronet sources from the 1990s as well as new observations taken
in 2005.  We finish this section with more detailed information about
a few particular sources.

\subsection{The Herbig Ae Star R CrA}

The Herbig Ae Star R CrA was first detected as a radio source in 1997,
reported by \citet{fcw98}. Subsequently, R~CrA was detected as a weak
radio source also by \citet{for05}, with a rather constant average
flux density of 0.23~mJy (the same as in 1997) in nine observations
carried out in summer 1998. However, R~CrA remained undetected at
radio wavelengths several times before: in September 1985 (VLA, 6~cm,
3$\sigma$=0.17~mJy, \citealp{bro87}), in February 1990 (VLA, 3.6~cm,
3$\sigma$=0.10~mJy, \citealp{ski93b}) as well as with the VLA and the
AT in 1985-87 and 1992, respectively \citep{sut96}. In August 2005,
during a multi-wavelength observing campaign, R~CrA was observed at
the highest flux density yet, 0.30~mJy (VLA, 3.6~cm \citealp{for06}),
indicative of long-term variability. The emission may be due to a
close companion to R~CrA and not to this Herbig Ae star itself (see
also Sect.~5).

\subsection{The Multiple System TY CrA}

The quadruple system TY CrA was first detected in centimetric radio
emission (6~cm) by \citet{bro87}. \citet{ski90} observed the source at
wavelengths of 3.6~cm and 6~cm, determining a negative spectral index
that would imply a nonthermal emission process, the only such case in
a sample of 57 Herbig Ae/Be stars \citep{ski93}. However, circular
polarization was below detectable levels in all observations.

\subsection{The Class~I Protostar IRS 5}

The Class I protostar IRS~5 was discovered to be the first -- and
until now only known, maybe apart from T Tau S -- Class~I protostar
showing circularly polarized nonthermal radio emission \citep{fcw98},
most likely gyrosynchrotron radiation. What makes this source so
different from the surrounding similar but unpolarized Class~I
protostars is unclear.  In the long-term variability analysis of
\citet{sut96}, IRS~5 was flaring once (there is no polarization
information) and \citet{fcw98} noted fast variability in the
polarization degree towards this source.  It also turns out to be the
most variable radio source in the \textsl{Coronet} cluster on
timescales of months \citep{for05}, showing several flares.  IRS 5 is
a close near-infrared binary \citep{che93,nis05}
as well as a radio-wavelength binary \citep{choi08} and it has
also been resolved in X-rays  \citep{fpm07,hamaguchi08}.
It is surrounded by a near-infrared
reflection nebula \citep{cas87}.

\subsection{The Protostars in IRS 7}

\begin{figure}
\centering
\includegraphics[width=7cm]{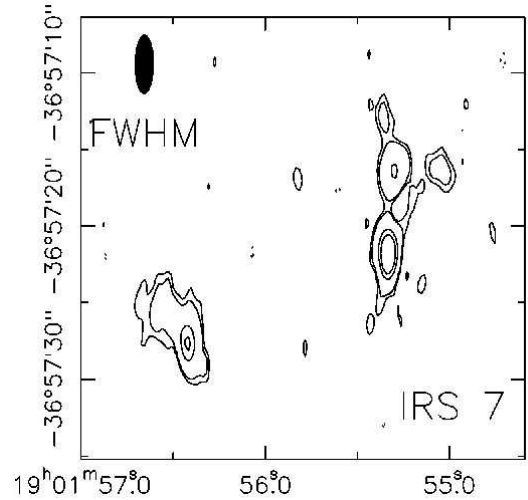}
\caption{IRS~7 in the 3.6~cm continuum, as seen with the NRAO VLA, contour lines delineate levels of 3, 5, 50 and 100 $\sigma$ with an rms noise level of $\sim12\mu$Jy (from \citealp{for05}). The eastern source is IRS~7B, the western source is IRS~7A with Radio Source 9 to the north of it.}
\label{for05irs7}
\end{figure}

The protostars around IRS 7 remain largely enigmatic. IRS~7 has several components, and it may either be a multiple protostellar system or we may see different
components in a jet: In centimetric radio emission (see Fig.~\ref{for05irs7}), the two brightest
sources in the \textsl{Coronet} cluster are close to IRS~7, i.e. IRS~7A (or 7W) and IRS~7B (or 7E),
the latter actually being the brightest one (and at the position
of infrared source IRS~7). There is an additional
radio source five arcseconds north of IRS 7A (Radio Source 9,
as defined by \citealp{bro87}), and three more, weak sources have been detected
in the surroundings \citep{for05}. Based on single-dish submillimeter data,
\citet{nut05} concluded that IRS~7A is an infrared-detected Class~I protostar while
IRS~7B appeared to be a Class~0 protostar.
However, \citet{pon03} detected both sources,
IRS~7A/B, at 4.8~$\mu$m using VLT-ISAAC, arguing that IRS~7B
remained undetected at 10~$\mu$m \citep{wil97} due to silicate absorption.
\citet{cho04} analyzed the region at 6.9~mm with the VLA.
While IRS~7B was only marginally detected, they found emission extended northwards from
IRS~7A, towards Radio Source 9. \citet{har01}, using ATCA at 6~cm,
found structure resembling a clumpy bipolar outflow north and south of IRS~7B.
They found that IRS~7A and IRS~7B may actually form a wide proto-binary surrounded
by a common rotating molecular torus. \citet{gro07} used high-resolution submillimeter observations in a multi-wavelength study
including also \textsl{Spitzer} mid-infrared data
and find two submillimeter continuum sources, coincident with IRS~7B (SMA~1) and Radio Source 9 (SMA~2). They concluded that
SMA~2 is a low-mass ($\leq0.3~M_\odot$) Class 0 source, while SMA 1 is a transitional Class 0/Class I object of higher envelope
mass ($\sim1-5~M_\odot$). The different evolutionary stages are difficult to reconcile with the idea that they form a protobinary.

\section{X-ray Observations}
\label{sec_xray}

First X-ray observations of the CrA region with the {\em Einstein} observatory
resulted in eleven detections of PMS stars \citep{wal86,wal97}, mostly wTTSs.
Optical follow-up observations of {\em Einstein} and ROSAT sources also led to the
identification of many new wTTSs and a few new cTTSs in and around CrA, see Sect. 2.

With {\em Einstein} and ROSAT, R CrA was searched for X-ray emission,
but to no avail \citep{zip94,dam94}.
The \textsl{Coronet} cluster started protostellar X-ray astronomy with the discovery of hard
X-rays from Class~I protostars with ASCA \citep{koy96}.
The angular resolution was still low (FWHM $\sim 2'$), so
that an enormous X-ray flare occuring during the observation could not be
clearly attributed to any one of the sources.
The X-ray emission of three Class~I protostars in the \textsl{Coronet} cluster
was later confirmed by \citet{nep97}, using ROSAT, see Table~\ref{ysotable}.
First results from comprehensive \textsl{Chandra}-ACIS imaging observations
of this region were presented by \citet{gar03} who detected more than 70 sources.
\citet{ham05} observed an enormous X-ray flare towards the Class~0/I source
IRS~7B, using XMM-\textsl{Newton}.
Multiple epochs of archival X-ray data were analyzed by \citet{for05} for
X-ray variability of radio-detected sources in the \textsl{Coronet} cluster.
Among these sources, IRS~5 appears to be the most variable one
(which it is also at radio wavelengths). Additionally, emission of surprisingly
hot plasma (100~MK) from R~CrA was found, possibly due to a very
close companion source (separated by only 10-15 AU) for which there
is some spectro-photometric evidence \citep{tak03}.
The quadruple PMS system TY CrA was already detected
by {\em Einstein} and ROSAT \citep{zip94,dam94}.
Long-term ASCA observations are discussed by \citet{ham05b} who observe
a flare with a long rise time, possibly inter-binary interaction.
While previously undetected, the Herbig Ae/Be star T CrA was
marginally detected in X-rays by \citet{ski04}, using \textsl{Chandra}.
The currently deepest X-ray dataset of the \textsl{Coronet} region is discussed
by \citet{fpr06}; 46 sources can be identified with infrared and/or optical sources.

\acknowledgements We would like to thank Bo Reipurth, Fernando Comer\'on,
Karl Menten, and John Graham for comments on earlier versions of this paper,
as well as Loke Kun Tan for the permission to reproduce his color photo of the CrA region.


\end{document}